\documentclass[journal=jacsat,manuscript=article]{achemso}

\usepackage[version=3]{mhchem} 
\usepackage{algorithm}
\usepackage{algpseudocode}

\author{Sooyong Chae}
\affiliation{LPICM, CNRS, Ecole polytechnique, IP Paris, Palaiseau, France}
\author{Tongyu Huang}
\affiliation{Graduate School at Shenzhen Tsinghua University, 
Shenzhen, China}
\author{Omar Rodr\'iguez-Nu\~nez}
\altaffiliation{Currently with Department of Neurosurgery, Inselspital, University of Bern, Switzerland}
\author{Th\'eotim Lucas}
\author{Jean-Charles Vanel}
\author{J\'er\'emy Vizet}
\altaffiliation{Currently with ArianeGroup, Mureaux, France}
\author{Angelo Pierangelo}
\affiliation{LPICM, CNRS, Ecole polytechnique, IP Paris, Palaiseau, France}
\author{Gennadii Piavchenko}
\affiliation{I.M. Sechenov First Moscow State Medical University, 
Moscow, Russia}
\author{Tsanislava Genova}
\affiliation{Biophotonics Laboratory, Institute of Electronics BAS, Sofia, Bulgaria}
\author{Ajmal Ajmal}
\author{Jessica C. Ramella-Roman}
\affiliation{Department of Biomedical Engineering, 
Florida International University, Miami, USA}
\alsoaffiliation{
Herbert Wertheim College of Medicine, Florida International University, Miami, USA.}
\author{Alexander Doronin}
\affiliation{School of Computer Sciences, Victoria University, Wellington, New Zealand}
\author{Hui Ma}
\affiliation{Graduate School at Shenzhen Tsinghua University, 
Shenzhen, China}
\alsoaffiliation{Department of Physics, Tsinghua University, Beijing, China}
\author{Tatiana Novikova}
\affiliation{LPICM, CNRS, Ecole polytechnique, IP Paris, Palaiseau, France}
\alsoaffiliation{Department of Biomedical Engineering, 
Florida International University, Miami, USA}
\email{tatiana.novikova@polytechnique.edu}

\title[An \textsf{achemso} demo]
  {Machine Learning Model for  Complete Reconstruction of  Diagnostic Polarimetric Images from partial Mueller polarimetry data\footnote{Sooyong Chae and Tongyu Huang are co-first authors}
   }

\abbreviations{IR,NMR,UV}

\begin{document}







\newpage
\begin{abstract}
The translation of imaging Mueller polarimetry to clinical practice is often hindered by large footprint and relatively slow acquisition speed of the existing instruments. Using polarization-sensitive camera as a detector may reduce instrument dimensions and allow data streaming at video rate. However, only the first three rows of a complete 4$\times$4 Mueller matrix can be measured. To overcome this hurdle we developed a machine learning approach using sequential neural network algorithm for the reconstruction of missing elements of a Mueller matrix from the measured elements of the first three rows. The algorithm was trained and tested on the dataset of polarimetric images of various excised human tissues (uterine cervix, colon, skin, brain) acquired with two different imaging Mueller polarimeters operating in either reflection (wide-field imaging system) or transmission (microscope)  configurations at different wavelengths of 550 nm and 385 nm, respectively. The reconstruction performance was evaluated using various error metrics, all of which confirmed low error values. The execution time of the trained neural network algorithm was about 300 microseconds for a single image pixel. 
It suggests that a machine learning approach with parallel processing of all image pixels combined with the partial Mueller polarimeter operating at video rate can effectively substitute for the complete Mueller polarimeter and produce accurate maps of depolarization, linear retardance and orientation of the optical axis of biological tissues, which can be used for medical diagnosis in clinical settings.  
\end{abstract}

\section{Introduction}
Imaging Mueller polarimetry has already demonstrated its potential in medical diagnosis.$^{1-14}$
Despite the promising results, the 
translation of this modality to clinical practice for real-time applications is challenging
due to limitations related to the physical setup of imaging devices and the speed of data acquisition and post-processing, which currently hinder the feasibility of \emph{in vivo} polarimetric imaging for diagnostic purposes.

For the complete  Mueller polarimeters, at least 16 measurements are required to get all elements of 4$\times$4 real-valued Mueller Matrix (MM) \cite{Goldstein, Azzam2016, ChipmanBook}. Performing the measurements sequentially makes the design and implementation of imaging Mueller polarimetric system simpler, but also less time-efficient  \cite{Novikova2016, 
Pardo2024, Gil2023, Huyhn2021, Meglinski}. 
The systems with spectral polarization coding can be very compact and measure a  MM at the kHz rate. However, taking wide-field polarimetric images with such systems will require its implementation in a scanning mode \cite{LeG2021, Dubr2012}. The use of photoelastic modulators for fast wide-field MM imaging was reported in~\cite{Gribble2016} , but such systems are quite bulky. The elegant non-conventional solution for single-shot acquisition of the complete MM based on using metasurfaces for polarization generation and analysis was recently suggested by A.~Zaidi et al.\cite{Zaidi2024} , but this approach is still in early development stage and will require
additional studies to translate it into the clinic.

Alternatively, the use of commercially available polarization-sensitive camera 
as the detector \cite{DoFP1, qi2017real, Got2021} may significantly reduce the instrument footprint and also the acquisition time due to the division of focal plane (DoFP) super-pixel arrays fitted with linear polarizers. 
Currently polarization-sensitive camera can only image linear polarization states, producing a reduced 3$\times$4 MM with fourth row missing. The design and implementation of the fast complete Mueller microscope making use of two polarization sensitive cameras combined with a beam splitter and quarter waveplate is described in\cite{Huang2021} , but such hardware solution increases both the footprint and the cost of an instrument. Algebraic solutions have been proposed, but they focus on the extraction of polarimetric parameters of a sample (diattenuation, retardance and depolarization) rather than reconstruction of its complete MM. Moreover, these algebraic solutions have their own drawbacks, e. g. time-consuming numerical solution of the system of non-linear equations \cite{DoFP2} or providing the incomplete set of polarimetric parameters \cite{novikova2022complete}. Thus, the reconstruction of the fourth missing row of MM from the measurements with a single polarization-sensitive camera setup, still remains a challenge. 

To bridge this gap, our studies leverage machine learning models to develop a method for accurately predicting the missing MM elements. The aim is to 
perform a reliable reconstruction of the last row of the complete 4$\times$4 MM using the information from the measured reduced 3$\times$4 MM. 

The paper is composed as follows: 
first, we introduce the basics of Stokes-Mueller formalism for polarized (or partially polarized) light and discuss the decomposition of MM, which 
generates the maps of depolarization, retardance and diattenuation of a sample. The experimental dataset of complete MM images acquired with the custom-built wide-field reflection imaging MM polarimeter (IMP1) using white light source on formalin-fixed thick human cervical specimens and with the transmission MM polarimetric microscope (IMP2) 
using UV-A (385 nm) light source
on thin histological sections of human skin, brain and colon tissues as well as the physical realizability filtering of experimental MMs is also described.
Next section is dedicated to the machine learning approach including the description of the objective, dataset splitting strategy, model selection and hyperparameter tuning. Then the results and discussion on application of the developed sequential neural network model to the experimental dataset of polarimetric images are presented and followed by the conclusions.

\section{Background}
\label{sect:background}
\subsection{Stokes Vector and Mueller Matrix}

The Stokes-Mueller formalism \cite{Goldstein,ChipmanBook} represents the theoretical framework that describes the polarization state of a light beam using a real-valued $4 \times 1$ vector, referred to as a Stokes vector $\mathbf{S}$ and defined as
\begin{equation}
\label{Eq:1}
\mathbf{S}=
\begin{pmatrix}
S_1 \\  S_2 \\ S_3 \\ S_4
\end{pmatrix} =
\begin{pmatrix}
I_{0^\circ}+I_{90^\circ} \\ I_{0^\circ}-I_{90^\circ} \\ I_{+45^\circ}-I_{-45^\circ} \\ I_{R}-I_{L}
\end{pmatrix}
\end{equation}
where $I_{0^\circ}$ and $I_{90^\circ}$ are the intensities of light measured after the beam passes through a linear polarizer oriented horizontally and vertically, respectively. $I_{+45^\circ}$ and $I_{-45^\circ}$ represent the corresponding intensities for a linear polarizer placed at the angles $+45^\circ$ and $-45^\circ$, $I_{R}$ and $I_{L}$ are the intensities of right- and left-handed circularly polarized light.
The degree of polarization $\rho$ of Stokes vector $\mathbf{S}$ is defined as
\begin{equation}
\label{Eq:4}
\rho=\frac{\sqrt{S_2^2+S_3^2+S_4^2}}{S_1}, \quad (0 \leq \rho \leq 1)
\end{equation}
\noindent with parameter $\rho$ equal to 0 for fully depolarized light, and equal to 1 for completely polarized light.
The following matrix equation describes the interaction of a light beam with any linear optical system: 
%
\begin{equation}
\label{eq:MuellerMatrix}
\begin{pmatrix}
S_1^{\text{out}} \\
S_2^{\text{out}} \\
S_3^{\text{out}} \\
S_4^{\text{out}}
\end{pmatrix}
=
\begin{pmatrix}
M_{11} & M_{12} & M_{13} & M_{14} \\
M_{21} & M_{22} & M_{23} & M_{24} \\
M_{31} & M_{32} & M_{33} & M_{34} \\
M_{41} & M_{42} & M_{43} & M_{44}
\end{pmatrix}
\begin{pmatrix}
S_1^{\text{in}} \\
S_2^{\text{in}} \\
S_3^{\text{in}} \\
S_4^{\text{in}}
\end{pmatrix}
\end{equation}

The 4$\times$4 real-valued transfer matrix $\mathbf{M}$ in Eq. \ref{eq:MuellerMatrix} is called Mueller matrix of a sample, 
and it describes the transformation of input Stokes vector $\mathbf{S}^{\text{in}}$ into output Stokes vector $\mathbf{S}^{\text{out}}$ upon interaction with
a sample. The Stokes-Mueller formalism, which
describes both fully or partially polarized (and even completely depolarized) light, is particularly suited for the description of light interaction with depolarizing samples, like biological tissues.

\subsection
{
Decomposition of Mueller Matrices}
\label{subsect:lu_chipman}
 The MM contains information on all polarimetric properties of a sample. 
However, the straightforward interpretation of the physical meaning of MM element values 
is possible for the MM elements of the first row and column only, which represent the diattenuation and polarizance properties of a sample. To understand the physical meaning of 
the values of other MM elements, the MM matrix decomposition approach is adopted 
yielding the values of
depolarization and retardation of a sample calculated
from its complete MM.
It was demonstrated by Lu and Chipman \cite{lu1996interpretation} that any physically realizable MM \cite{cloude1986group, cloude1990conditions} can be expressed as the product of corresponding MMs of a diattenuator \((\textbf{M}_D\)), a retarder ($\textbf{M}_R$), and a depolarizer ($\textbf{M}_{\Delta}$) \cite{Goldstein, ChipmanChapter}:
\begin{equation}
\label{eq:Lu_chipman}
\textbf{M} = \textbf{M}_{\Delta}\textbf{M}_R\textbf{M}_D.
\end{equation}

Applying such decomposition pixel-wise one can obtain the maps of depolarization, scalar linear retardance, and orientation of the optical axis, which may increase the contrast between healthy and pathological zones of tissue and improve medical diagnosis.$^{10,37,38}$ 
The original version of Lu-Chipman decomposition cannot be used on a reduced 3$\times$4 MM. Several attempts were made to develop the decomposition algorithms for such reduced matrix in order to extract the relevant polarimetric parameters of a depolarizing sample. However, some of these algorithms either require the numerical solution of the system of non-linear equations pixel-wise \cite{DoFP2} or provide not a complete set of polarimetric parameters \cite{novikova2022complete}. We suggest overcoming these limitations by using a machine learning approach for the reconstruction of the missing fourth row of MM. It will allow us to combine the advantages of using partial Mueller polarimetry for the fast acquisition and the existing algorithm of the decomposition of complete MMs. 

\section{Dataset}
\label{sect:dataset}
The primary objective of our studies is to uncover hidden patterns within the first three rows of MM images of biological tissues and link them to the patterns of the last row of MM images by building the neural network (NN) model. 
\subsection{Samples of biological tissues}
Our dataset is composed of complete MM images of various biological samples, namely, formalin-fixed bulk human cervical tissues, thin sections of human brain stained with silver pre-treatment and impregnation
\cite{Piav2022}, thin sections of skin and colon stained with hematoxylin \& eosin (H\&E) \cite{HE}, which were 
measured with two different imaging Mueller polarimetric systems. 

\begin{figure}[h!]
    \begin{center}
    \includegraphics[width=.37\columnwidth]{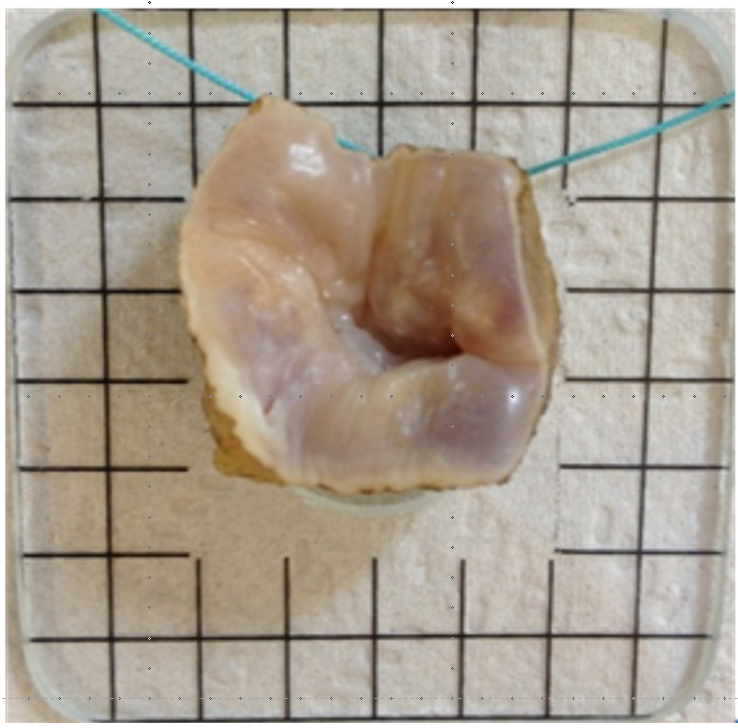}
    \quad
    \includegraphics[width=.42\columnwidth]{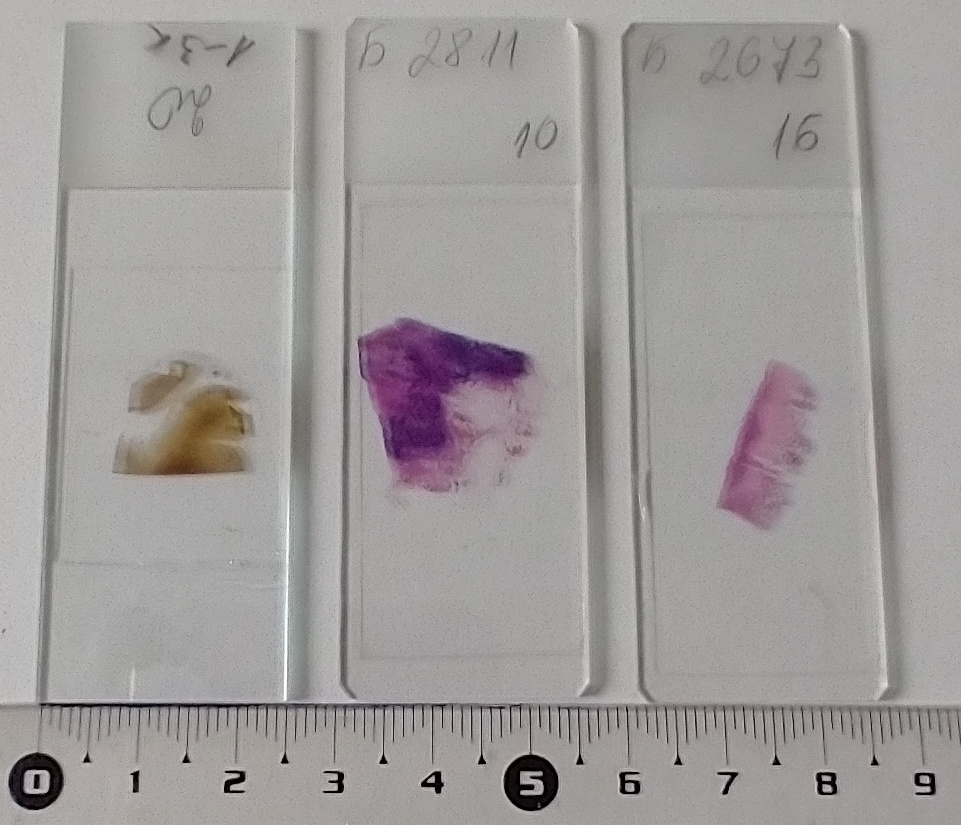}
    \end{center}
    \caption{Photos: (left) formalin-fixed cervical specimen (adapted from$^{10}$
    ). The grid engraved into the glass sample holder indicates 5 mm $\times$ 5 mm scale; (right) glass microscopy slides with stained sections of 
    human brain (5 $\mu$m thick), 
    human colon with cancerous polyp (10 $\mu$m thick), and 
    human skin with basal cell carcinoma (16 $\mu$m thick).}
    \label{Photos}
\end{figure}

Informed consent was obtained from all patients. The approvals were obtained from the local ethics committees of the University Hospital "Tsaritsa Yoanna – ISUL", Sofia, Bulgaria (Ref. \#286/2012), I.M. Sechenov First
Moscow State Medical University, Moscow, Russia (protocol \#03-19, March 2019) and Institut National du Cancer and Canc\'erop$\hat{\text{o}}$le, France (PAIR
Gyn\'eco contract \#2012-1-GYN-01-EP-1). 
%
The photos of several biological tissue samples
are shown in Fig. \ref{Photos}. 
%
\subsection{Measurements with the IMP1 system}
The first part of the dataset consists of wide-field MM images obtained from formalin-fixed conization specimens of uterine cervices from 23 patients. These measurements were conducted using a liquid crystal-based imaging MM polarimeter (IMP1) operating in a reflection mode within the visible wavelength range. Measurement accuracy was ensured by implementing the eigenvalue calibration method.$^{41}$ 
Details of the data acquisition protocol are available in D. Robinson et al.$^{10}$ 
and a comprehensive description of the optical layout of the IMP1 system is provided in the Supplementary Materials. 
Specimens were collected from 23 patients at the Kremlin-Bicetre University Hospital, France, all of whom had histologically confirmed high-grade cervical intraepithelial neoplasia (CIN3 or precancer). 
Fig. \ref{fig:Non-filtered}a shows an example of MM images of bulk formalin-fixed cervical specimen measured with the IMP1.
\subsection{Measurements with the IMP2 system}
The second part of the dataset consists of images acquired using the custom-built transmission MM microscope (IMP2). 
This system features two continuous light sources: a UV-A LED (385 nm) and a white light LED. 
It includes a Polarization State Generator (PSG) with a fixed Glan-Taylor polarizer, a rotating achromatic quarter-wave plate, and a Polarization State Analyzer (PSA) composed of the same components arranged in a reverse order. The IMP2 makes use of two microscope objectives (4$\times$ and 20$\times$) and a CMOS camera for the detection.  
\begin{figure*}[ht]
\label{Fig:1}
\centering
    \centering    \includegraphics[width=0.48\linewidth]{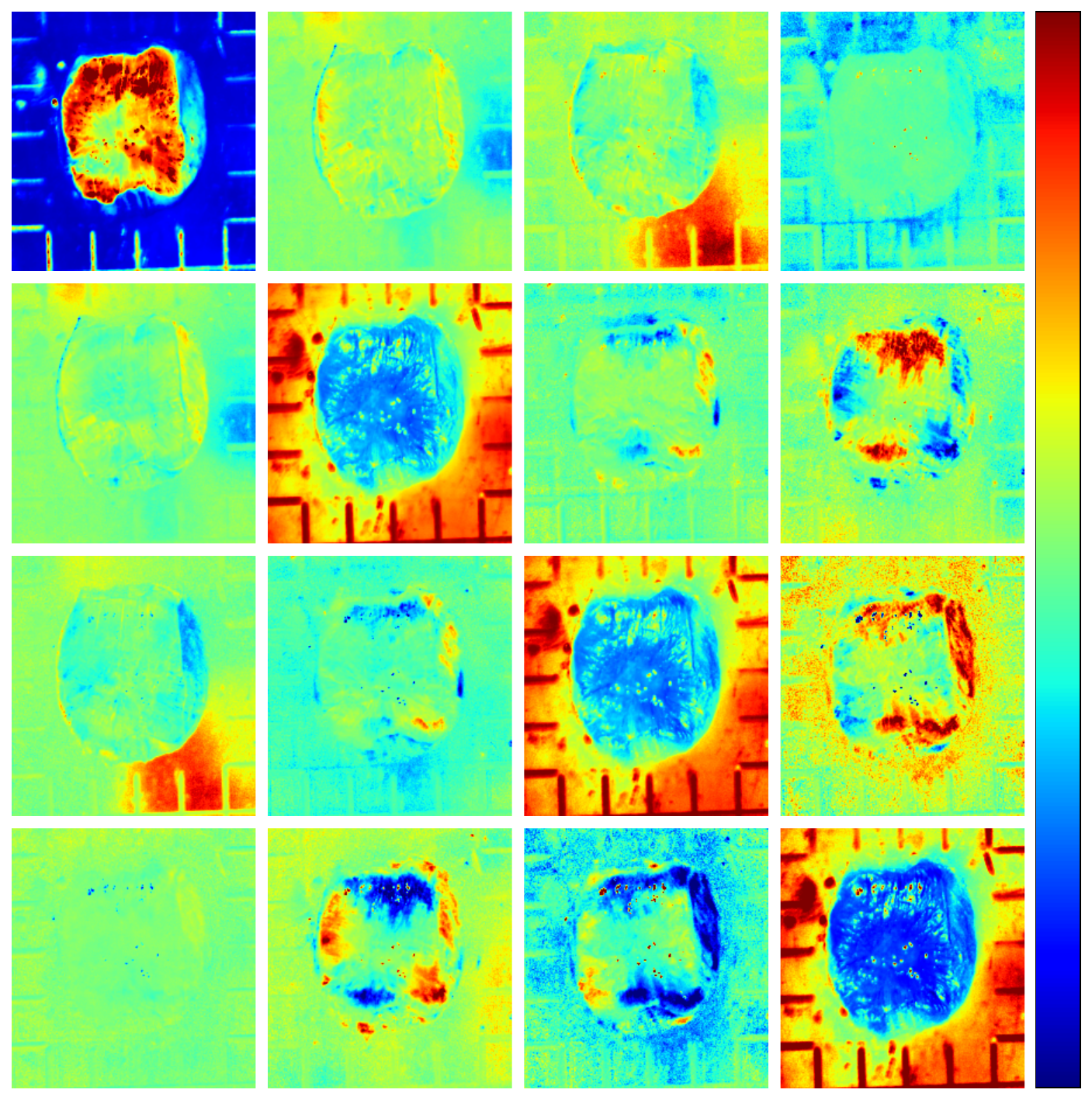}
\hfill
    \centering    \includegraphics[width=.48\linewidth]{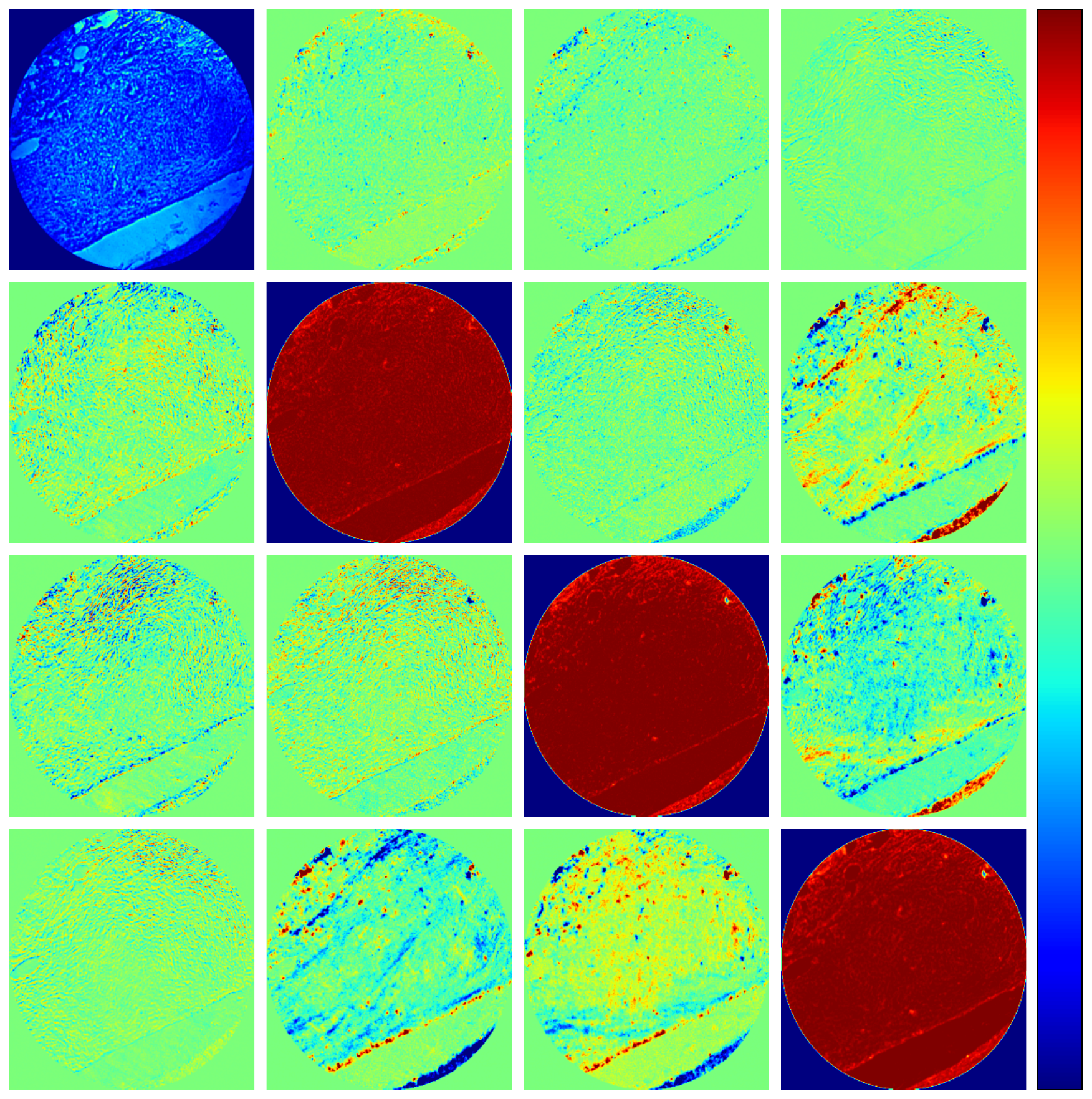}
%
\caption{Examples of non-filtered (see Sec. \ref{sect:physically_realizability_test}) MM images of a) bulk cervical tissue
recorded with IMP1; b) 16 $\mu$m thick section of healthy human skin tissue recorded with IMP2, the field of view 
is 1 mm.
The color bar for the diagonal elements ranges from 0 to 1, for the off-diagonal elements from -0.1 to 0.1. \(M_{11}\) shows raw intensity values, all other elements of MM are normalized by \(M_{11}\) values pixel-wise.}
\label{fig:Non-filtered}
\end{figure*}
%

The IMP2 acquires 16 intensity images by sequentially rotating both waveplates to four pre-selected angles$^{42,43}$ of the optical axis $\pm$51.7$^\circ$, $\pm$15.1$^\circ$ 
for the following reconstruction of MM. A hybrid calibration method was implemented to 
eliminate the influence of instrument systematical errors.$^{44}$ 
The detailed description and optical layout of the IMP2 system is provided in the 
Supplementary Materials.

Using the IMP2 system with UV-A source in order to account for data spectral variability, thin tissue sections 
were measured in transmission, including human brain (5 various samples, 26 regions including white and gray matter zones), human skin (3 samples, 15 regions including healthy and basal cell carcinoma zones) and human colon (3 samples, 11 regions including healthy and cancerous zones). Hence, a total of 42 MMs 
have been collected on tissue sections from different patients and different pathological conditions of tissue for the second part of dataset.
Fig. \ref{fig:Non-filtered}b shows an example 
of MM images of healthy human skin section measured with IMP2. 

Tab. \ref{tab:1} details the comparative characteristics of the imaging MM systems and tissue types used in our studies for dataset acquisition.

\begin{table}
  \caption{Characteristics of IMP1 and IMP2 and measured samples}
  \begin{tabular}{lll}
    \hline
       & IMP1  & IMP2\\
    \hline
    Field of view       &   Up to 10 cm         &   Up to 1 mm                 \\
    Resolution (pixels) &  600 $\times$ 800     &  500 $\times$ 500            \\ 
    Imaging geometry    &  Reflection           &  Transmission                \\ 
    Wavelength          &  550 nm (visible)     &  385 nm (UV-A)               \\
    Tissue type         & Uterine cervix (bulk) & Brain, skin, colon (sections) \\
    Tissue thickness    & 1 cm - 3 cm           & 5 $\mu$m - 30 $\mu$m         \\
    \hline
  \end{tabular}
  \label{tab:1}
\end{table}


The dataset  collected with two different instruments on various types of tissues was employed to account for data variability and test the robustness and generalizability of the machine learning approach for reconstructing the elements of the fourth row of MM. 
\subsection{Data Processing: Physical Realizability Test}
\label{sect:physically_realizability_test}

Both IMP1 and IMP2 systems were calibrated to exclude the impact of the instrument's systematic errors. However, even small residual errors related to 
measurement noise may significantly affect the quality of polarimetric data. For example, physically non-realizable  MM may convert the Stokes vector of incoming light into the Stokes vector of outgoing light with $\rho>1$ (see Eq.~\ref{Eq:4}), which is wrong. Thus, before applying the machine learning approach to our experimental MM dataset for the reconstruction of the elements of the fourth row of MM we need to filter out all pixels with physically non-realizable MMs. 

There are various implementations of MM physical realizability test outlined in \cite{novikova2024time} .
We opt for the evaluation of the sign of four coefficients of characteristic polynomial (CCP) of coherency matrix \textbf{H} \cite{cloude1986group} using its expression via Pauli matrices \cite{Pauli} (see below).
\begin{algorithm}
\caption{Physical Realizability Test via Calculation of the Coefficients of Characteristic Polynomial of Coherency Matrix}
\label{alg:filter_matrices}
\begin{algorithmic}[1]
\State \textbf{Input:} Set of $4\times4$ matrices $\{\mathbf{M}_{k}, k=1,..., N\}$, 
$N$ is the total number of image pixels
\State \textbf{Output:} Filtered set of matrices $\{\mathbf{M}_{k}, k=1,..., N\}$, with 
a subset of $N_1$ physically realizable 
MMs
\Function{FilterMatrices}{dataset}
    \State Initialize an empty list \textit{filteredMatrices}
    \State Initialize $N_1 \gets 0$ \Comment{for physically realizable MMs}
    \For{each matrix $\mathbf{M}_{k}$ in \textit{dataset}}
        \State Compute CCP
        of the 
        coherency matrix $\mathbf{H}_{k}$ 
        \If{all coefficients are $\geq 0$}
            \State Add $\mathbf{M}_{k}$ to \textit{filteredMatrices}
            \State Increment $N_1 \gets N_1 + 1$
        \Else
            \State Replace all elements of $\mathbf{M}_{k}$ with 0
            \State Add the zero matrix $\mathbf{M}'_{k}$ to \textit{filteredMatrices}
        \EndIf
    \EndFor
    \State \Return \textit{filteredMatrices}
\EndFunction
\end{algorithmic}
\end{algorithm}

\begin{figure*}[ht]
\centering
    \centering
    \includegraphics[width=0.48\linewidth]{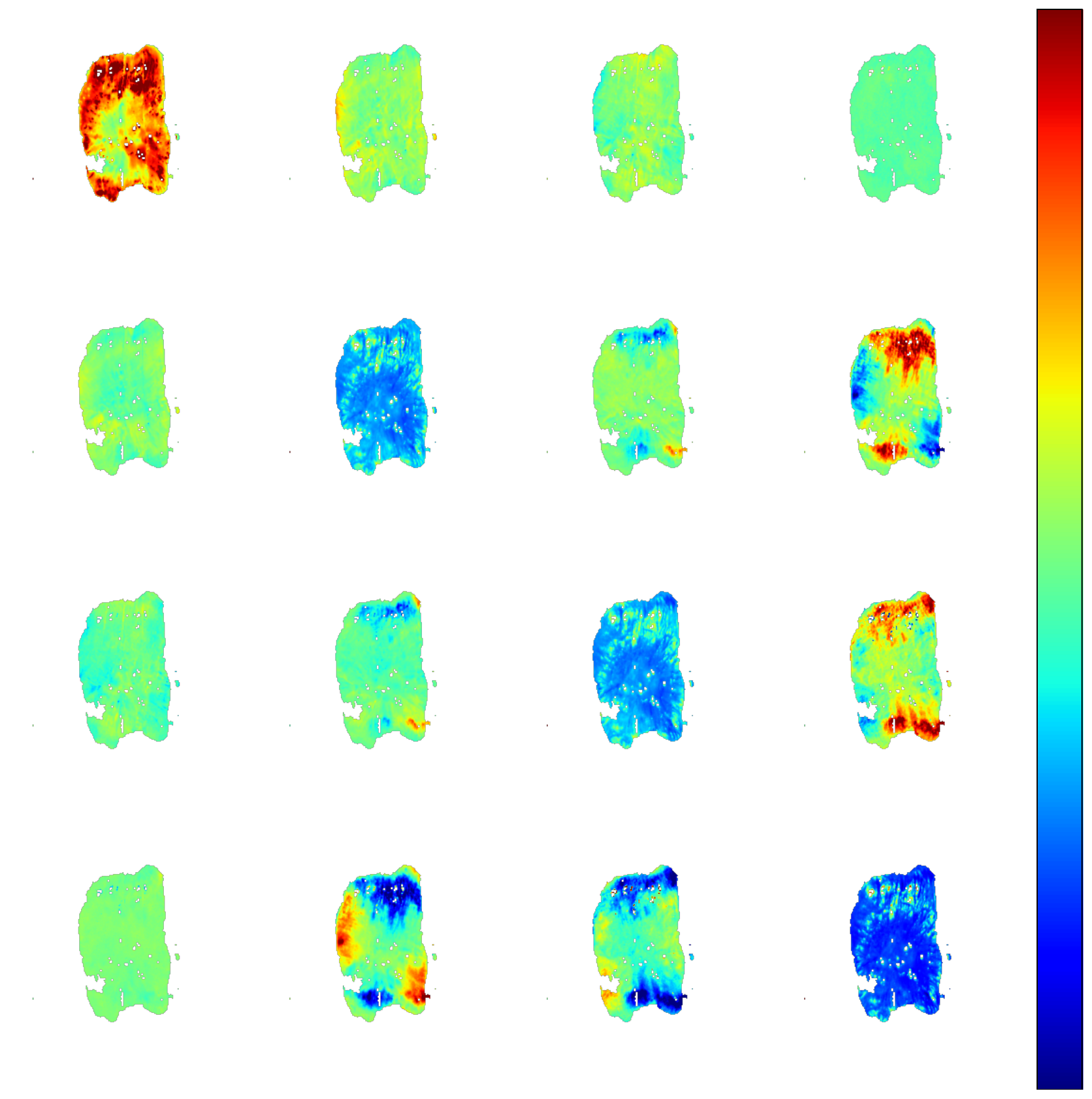}
\hfill
    \includegraphics[width=0.48\linewidth]{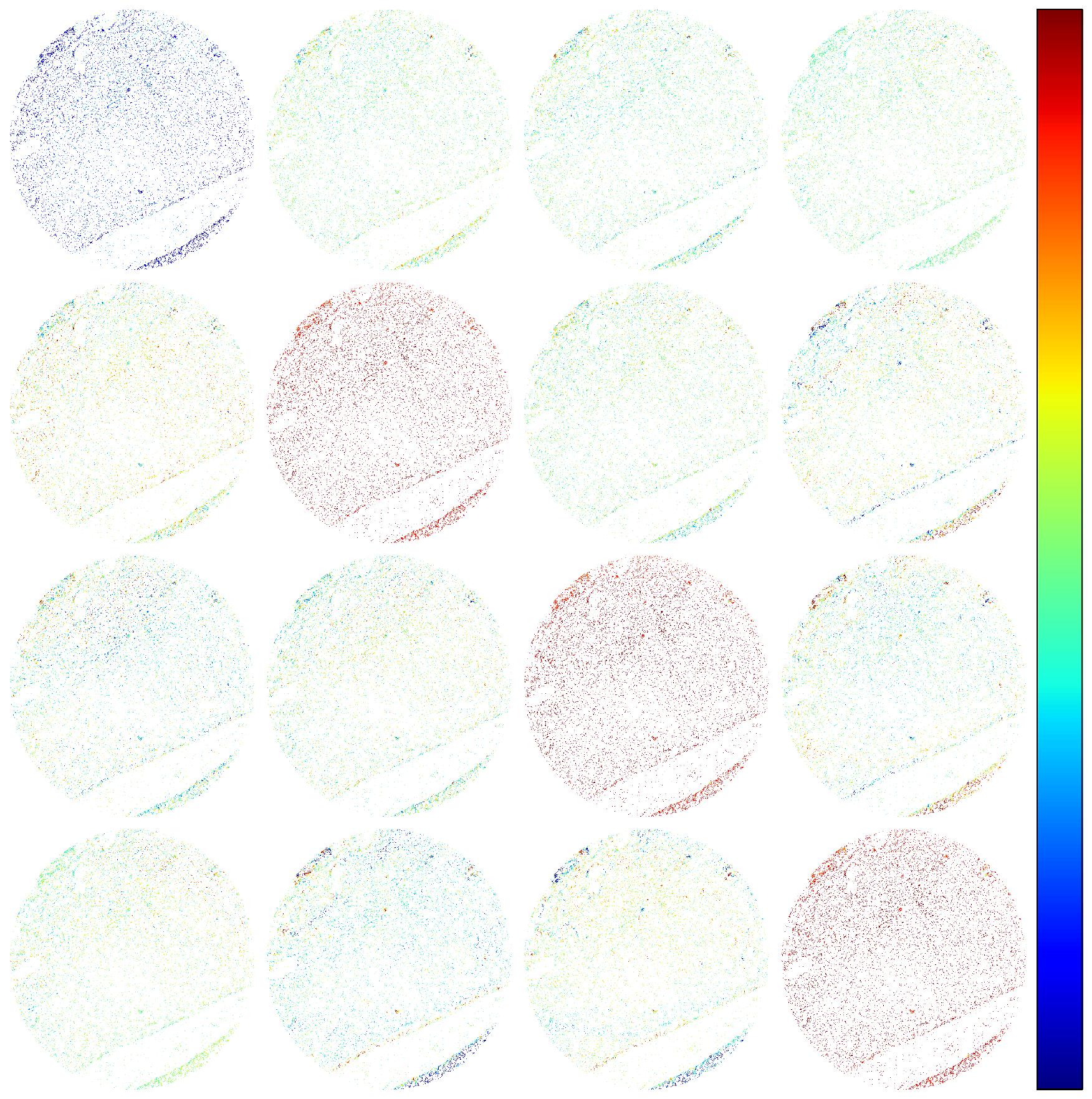}
\caption{Analogous to Fig. \ref{fig:Non-filtered}, the corresponding data were filtered by applying the 
Algorithm \ref{alg:filter_matrices}. The pixels with zero values, 
which indicate filtered out data, have been rendered in white
to enhance the visibility of 
zones of
physically realizable MMs.}
\label{fig:Filtered}
\end{figure*}

 This algorithm has a shorter execution time compared to widely used tests based on calculations of the eigenvalues of coherency matrix \textbf{H} (13.1 ms versus 561.3 ms, respectively, using 1 CPU core for the dataset of 420,000 MMs provided in \cite{novikova2024time}), because the former does not require the direct calculations of the elements of coherency matrix \textbf{H}. The results of the implementation of Algorithm \ref{alg:filter_matrices} for the datasets shown in Fig. \ref{fig:Non-filtered}a and Fig. \ref{fig:Non-filtered}b are represented in Fig. \ref{fig:Filtered}a and Fig. \ref{fig:Filtered}b, respectively.%
The zone of tissue in Fig.~\ref{fig:Filtered}a has been selectively retained by Algorithm \ref{alg:filter_matrices}, whereas the pixels of background and zones of specular reflection were filtered out and rendered in white in the images of cervical tissue acquired with the IMP1. 
The physical realizability test was also applied to the MM images acquired with the IMP2 resulting in the filtering out of the bare glass zones with no tissue (see Fig.~\ref{fig:Filtered}b).

\section{Machine Learning Approach}
Due to extremely rich information on tissue microstructure encoded in 16 images of MM, there are continuous attempts to combine imaging Mueller polarimetry with machine learning approach in order to improve the accuracy of biomedical diagnosis \cite{Nguyen2024, Kistenev2023, 
Sampaio2023, JRR, Pham2022, Moriconi}. Some of the classifiers use the compressed polarimetric data (maps of depolarization, retardance and diattenuation) as the input dataset \cite{LeeApplOpt}, others work with the input dataset composed of all MM images.$^{10,49}$ 

Currently, we do not target the diagnostic segmentation of the polarimetric images with machine learning algorithms. 
The primary objective of our studies is the development and implementation of a NN model for the accurate reconstruction of missing elements of experimentally measured 3$\times$4 MMs for the development of the fast-imaging Mueller polarimeter for real-time biomedical applications.
\subsection{Objective}
Given a partial 3$\times$4 MM, 
the vector \( \mathbf{x} \) is defined as the vector
form of the partial MM:

\begin{equation}
\mathbf{x} = \begin{bmatrix} M_{11}, M_{12},..., M_{33}, M_{34} 
\end{bmatrix} 
\end{equation}

The goal is to predict the elements of the last row \( \mathbf{y} \) of the complete MM: 
\begin{equation}
\mathbf{y} = \begin{bmatrix}
\hat{M}_{41}, \hat{M}_{42}, \hat{M}_{43}, \hat{M}_{44}
\end{bmatrix} 
\end{equation}

A model, denoted as \( f \), is used to estimate the last row \(
\mathbf{y}
\) based on the input vector \( \mathbf{x} \):

\begin{equation}
\mathbf{y}
= f(\mathbf{x})
\end{equation}

Upon deriving the predicted last row \( 
\mathbf{y}
\), the complete MM \( \textbf{M} \) is reconstructed to its complete \(4 \times 4\) form:

\begin{equation}
\textbf{M} = \begin{pmatrix}
M_{11} & M_{12} & M_{13} & M_{14} \\
M_{21} & M_{22} & M_{23} & M_{24} \\
M_{31} & M_{32} & M_{33} & M_{34} \\
\hat{M}_{41} & \hat{M}_{42} & \hat{M}_{43} & \hat{M}_{44}
\end{pmatrix}
\end{equation}

The model \( f \) encapsulates the machine learning model used to predict the last row elements $\hat{M}_{4i}$, ($i=$1,\dots,4)
and compared against the real dataset for the calculations of the performance metrics. In this study, two versions of the training model will be run.

The initial training
\textbf{Model I} will utilize only the first part of dataset collected with the IMP1 system on bulk cervical tissue specimens. Following this, the model robustness testing will be conducted using the data for isolated samples measured with either IMP1 or IMP2 system. This approach evaluates the model's robustness and identifies whether common characteristics exist between these two different parts of dataset. Subsequent
\textbf{Model II} will combine both 
parts of dataset to assess the model's generalizability for various biological tissues, different polarimetric instruments and measurement configurations used for data acquisition.
\subsection{Dataset Splitting Strategy}
There are different strategies for dataset splitting when using machine learning approach.$^{10,54}$ 
In our studies, before the machine learning model training 
step, the data from three samples were isolated from first part of training dataset measured with IMP1 system
for the consequent robustness testing. 
The selection of these testing samples was based on the variance of image quality and tissue structure. For each tissue type in the second part of dataset measured with the IMP2 system, the first sample was isolated for robustness testing. The selection of the sample area to be imaged with the IMP2 system was random. 
Therefore, no further random sampling technique was applied.

For data splitting 
a standard strategy was adopted\cite{Gholamy2018}, where the initial dataset was split into 80\% training set, 10\% of validation and 10\% of test set, respectively. The validation dataset was used during the model's training phase to fine-tune the hyperparameters of a model and to avoid model over-fitting. The test set was used for model evaluation after the training phase is completed to assess the model's performance metrics using Mean Squared Error (MSE), Mean Absolute Error (MAE), Root Mean Squared Error (RMSE) \cite{draper1998applied} and coefficient of determination \(R^2\) score \cite{Chicco2021}.
\subsection{Model Selection and Hyperparameter Tuning}
\begin{figure}[H]
\centering
\includegraphics[width=.5\columnwidth]{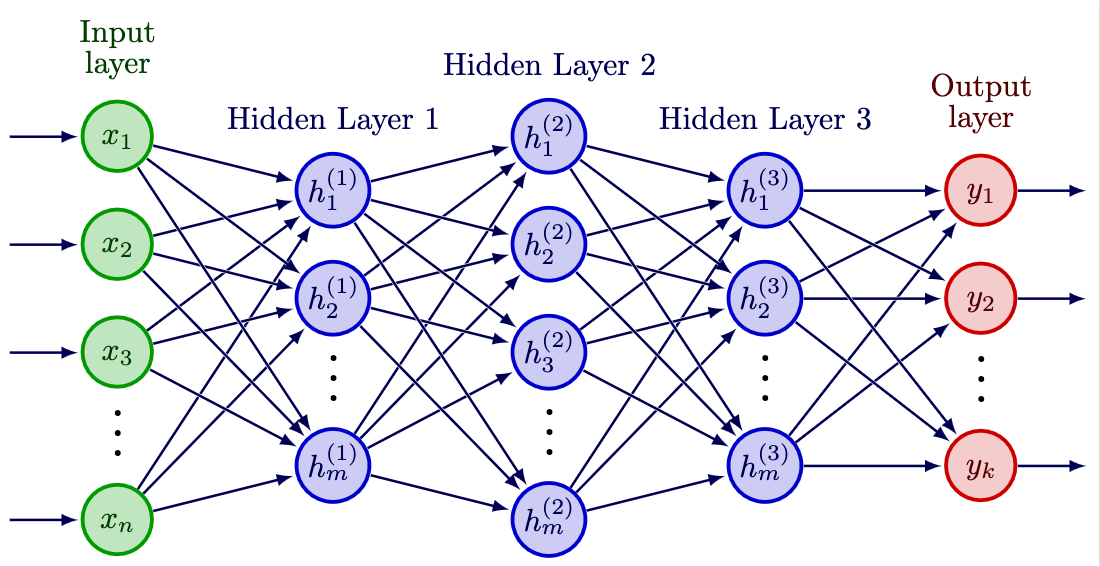}
\caption{High-level representation of sequential NN model architecture.}
\label{fig:nn_architecture}
\end{figure}

Prediction of the last row of the MM using machine learning algorithm is a novel approach so there is no established standard process to develop upon. Nevertheless, hidden patterns within the MM images are complex and require using advance learning techniques. Sequential NN model\cite{keras2024sequential} is adopted in this study due to the architecture's suitability for predicting continuous outputs, such as the components of MMs from multifaceted feature sets. This model architecture, with layers arranged in a linear stack, is capable of capturing complex nonlinear relationships inherent in high-dimensional data. Fig. \ref{fig:nn_architecture} provides visual representation of the NN architecture. 
The implementation of the NN model was executed using TensorFlow and Keras libraries\cite{tensorflow2015-whitepaper, chollet2015keras}. The hyperparameter search space was defined to explore various configurations of the NN architecture using Hyperband \cite{Li2018} algorithm. The algorithm finds the optimal combination of neurons in each layers, and the learning rate of the optimizer. The parameter search space was initially defined manually and adjusted when the search algorithm generated edge cases of the search space. The details of the final NN architecture are summarized in 
the Appendix.

\section{Results \& Discussions}
\label{sect:results}
%
\subsection{Model Training with the IMP1 data}
The training dataset for the first model contained MM images of cervical tissue only, and consisted of 80\% of 21 complete MMs of 16 images sized 600$\times$800 pixels, which were filtered using physical realizability test described above.
The MM images of the 3 remaining samples were isolated for further robustness test of the algorithm.

Thus, the first training algorithm used 2,158,584 matrices
and required 1,706.80 seconds to complete the training process. The application of the sequential NN model demonstrated strong predictive performance of the model in estimating the values of elements of the fourth row of MM. Most of the selected performance metrics underscore its effectiveness across different samples. The model achieved the following average performance metrics values across the four elements of last row of MM: 
\begin{itemize}
    \item 
    RMSE of 0.012166, 
    \item 
    MAE of 0.008848,
    \item 

    \(R^2\) of 0.828480,
\end{itemize} 
indicating high accuracy and robust predictive capability of the model.    

When inspecting individual performance scores for each element
, it is noted that despite having low error scores, there is underperformance in accuracy score \(R^2\) for the MM element $M_{41}$ (see Tab. \ref{tab:model_metrics}). 
\begin{table}[h]
\caption{\textbf{Model I}: Performance Metrics}
\label{tab:model_metrics}
\begin{tabular} {lllll} 
    \hline 
     \textbf{Metric} &  \textbf{M$_{41}$} & \textbf{M$_{42}$} & \textbf{M$_{43}$} & \textbf{M$_{44}$} \\
    \hline 
    MSE & 0.000032 & 0.000147 & 0.000147 & 0.000351 \\
    RMSE & 0.005681 & 0.012123 & 0.012129 & 0.018732 \\
    MAE & 0.004208 & 0.008541 & 0.009096 & 0.013547 \\
    $R^2$ & 0.483399 & 0.903402 & 0.937968 & 0.989152 \\
    \hline
\end{tabular}
\end{table}

\begin{figure}[!h]
\centerline{\includegraphics[width=.5\columnwidth]{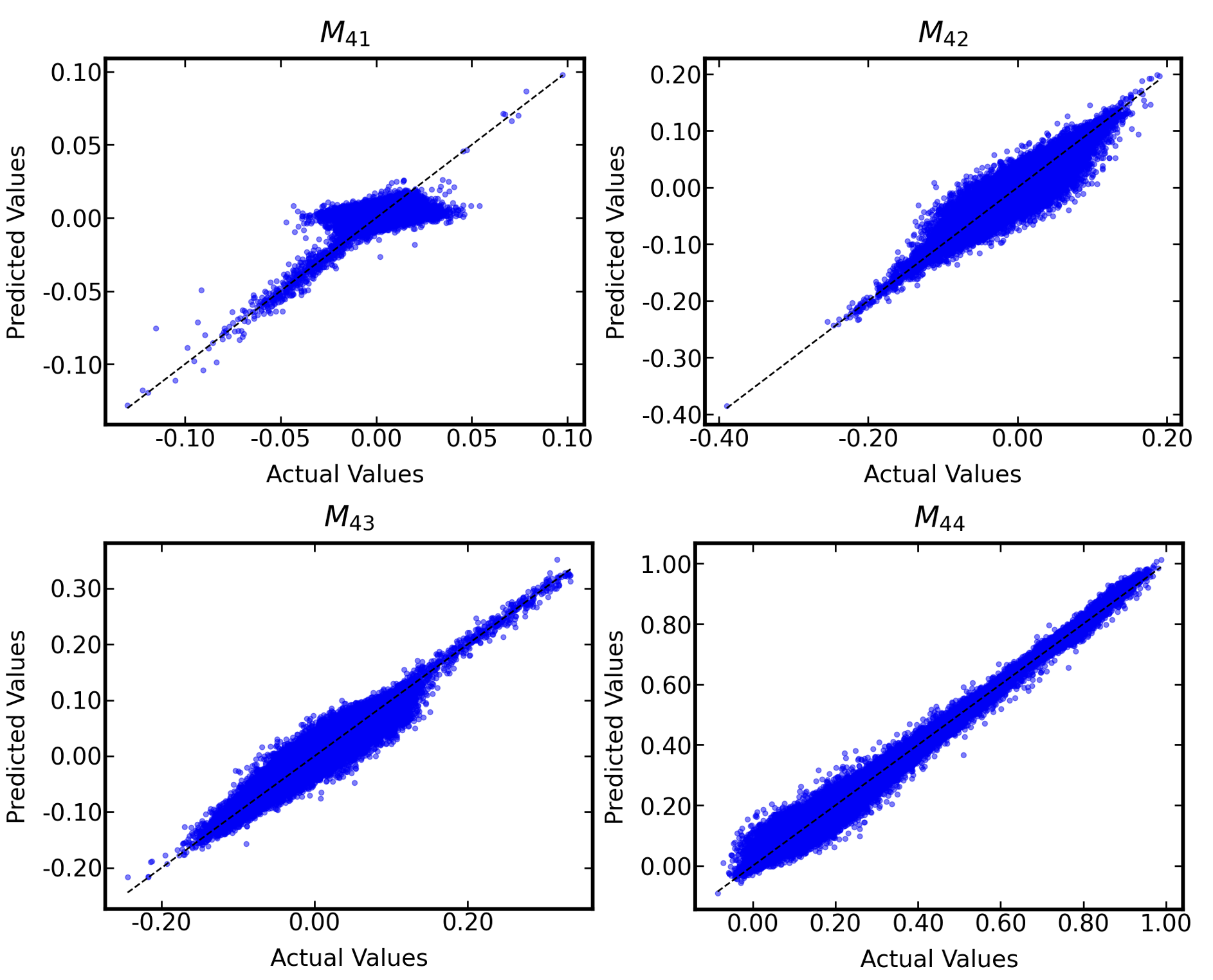}}
\caption{\textbf{Model I}: Scatter plots of predicted versus actual values of the last row elements of MMs from the test dataset measured with the IMP1.}
\label{fig:scatter actual_predicted_plot}
\end{figure}

Fig. \ref{fig:scatter actual_predicted_plot} illustrates the relationship between actual (X-axis) and predicted (Y-axis) values of the elements of fourth row of MMs.
%
This reveals a generally linear pattern for all four 
scatter plots that are closely aligned along the dotted line, which represents 
the prediction accuracy of an ideal NN model. Notably, the linear pattern for \(M_{41}\) seems to have a ceiling effect at a predicted value of zero. 
As was reported in \cite{Vitkin1} the diattenuation and polarizance of bulk biological tissues that are imaged at normal incidence in reflection configuration should be close to zero. Hence, the values of $M_{14}$ and $M_{41}$ elements of the corresponding MMs, which represent circular diattenuation and circular polarizance, respectively, should be close to zero as well. Our experimental data support it (see Fig. \ref{fig:Filtered}a). 
The random fluctuations of $M_{41}$ values in the vicinity of zero are related to the minor 
errors ($\sim$1-2\%) of the IMP1 measurements due to the instrument noise.
This observation suggests that the developed \textbf{Model I} effectively fits values of $M_{41}$ element, thus, 
performing as intended.
\subsubsection{Robustness Testing}
Using the isolated sample, robustness testing of the trained model was conducted. This step was carried out to ensure that the model is working as intended and to confirm there is no over-fitting of model parameters.
The first three rows of MM images of isolated sample were used as input into the trained model for evaluation of the accuracy of reconstruction of the elements of fourth row of MM. Table \ref{tab:sample25_model_metrics} presents the performance metrics for the isolated sample. 
\begin{table}[ht]
\caption{\textbf{Model I}: Robustness Testing using the IMP1 data}
\begin{tabular}{lllll} 
\hline
\textbf{Metric} & \textit{M$_{41}$} & \textit{M$_{42}$} & \textit{M$_{43}$} & \textit{M$_{44}$} \\
\hline
MSE & 0.000019 & 0.000092 & 0.000152 & 0.000465 \\
RMSE & 0.004344 & 0.009586 & 0.012324 & 0.021561 \\
MAE & 0.003358 & 0.007139 & 0.009434 & 0.014206 \\
$R^2$ & 0.539363 & 0.942836 & 0.910654 & 0.967196 \\
SSIM & 0.897079 & 0.939859 & 0.959785 & 0.992395 \\
\hline
\end{tabular} 
\label{tab:sample25_model_metrics} 
\end{table}

\begin{figure}[!h]
\centerline{\includegraphics[width=.5\columnwidth]{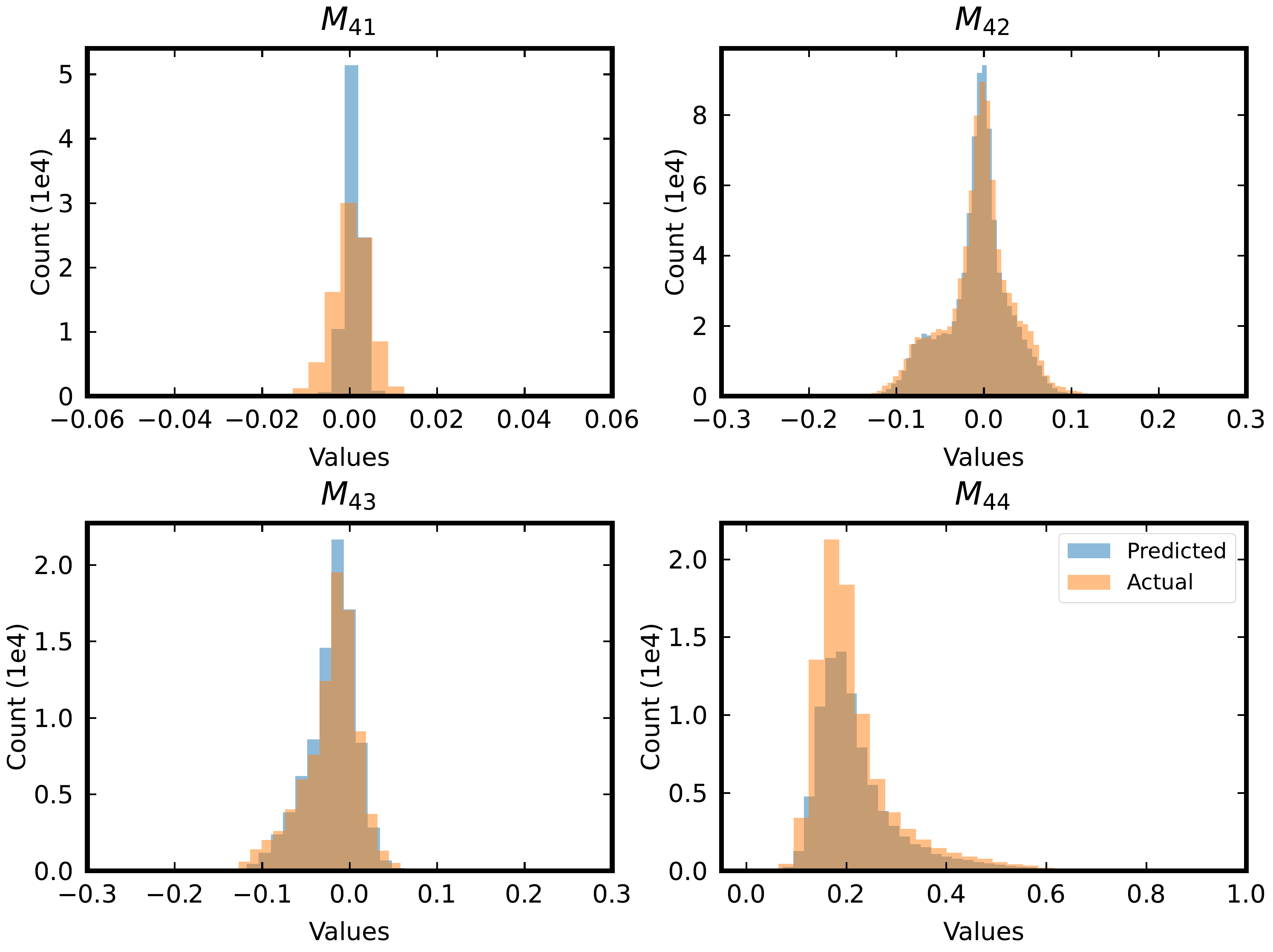}}
\caption{\textbf{Model I}: Histograms 
of actual and predicted values of the last row elements of MM 
for the isolated cervical sample measured with the IMP1. Only 
values for physically realizable MMs are shown.
}
\label{fig:histogram25_actual_predicted_plot}
\end{figure}

In addition to the standard metrics, the Structural Similarity Index (SSIM) \cite{wang2004image} has been calculated. SSIM is a perceptual metric used to evaluate the quality of images and videos. Unlike traditional metrics such as MSE, which primarily focuses on pixel-wise differences, SSIM assesses changes in structural information, luminance, and contrast, offering a more comprehensive evaluation of image similarity. The performance metrics in Tab. \ref{tab:sample25_model_metrics} confirm good accuracy on the predicted elements of MM
comparable to that shown in Tab.~\ref{tab:model_metrics}. 
Despite quite low \(R^{2}\) value for \(M_{41}\) (see the discussion above)
the large values of SSIM scores show the model's ability to maintain structural integrity and similarity in the predicted images.
Fig.~\ref{fig:histogram25_actual_predicted_plot} shows the histograms of the actual and predicted values of the last row elements of MM for one of the isolated samples. The general pattern appears consistent, with no noticeable differences. 

Table \ref{tab:afmmm_testing} presents the values of performance metrics of \textbf{Model I} tested on MM images of a skin sample recorded with the IMP2. 
\begin{table}[H]
\caption{\textbf{Model I}: Robustness Testing using the IMP2 data}
\label{tab:afmmm_testing}
\begin{tabular}{lllll}
\hline
\textbf{Metric} & \textit{M$_{41}$} & \textit{M$_{42}$} & \textit{M$_{43}$} & \textit{M$_{44}$} \\
\hline
MSE & 0.000077 & 0.000249 & 0.000363 & 0.000966 \\
RMSE & 0.008760 & 0.015766 & 0.019052 & 0.031078 \\
MAE & 0.003810 & 0.009327 & 0.009475 & 0.028786 \\
$R^2$ & 0.267674 & 0.292661 & -0.310504 & 0.993179 \\
SSIM & 0.773498 & 0.878399 & 0.794677 & 0.830971 \\
\hline
\end{tabular}
\end{table}

As expected, the performance is worsening. The error-based performance scores are generally acceptable, considering that the trained \textbf{Model I} was previously exposed only to the first part of dataset measured with the IMP1 system on cervical tissue at visible wavelength. However, the values of $R^2$ 
metric still indicate very low performance, particularly for \(M_{41}\), \(M_{42}\), and \(M_{43}\)
images. The discrepancy in the performance suggests inherent differences in optical patterns exist between various tissue types.
\subsection{Model Training with IMP1 and IMP2 Data Combined}
The trained model has so far been exposed solely to a dataset comprising MM images of cervical tissue, which raises important questions regarding its generalizability across different types of biological tissues. The training is extended to include both IMP1 and IMP2 MM images collected on various biological tissue types and different tissue zones to explore this aspect further.
%
The combined dataset provided 5,891,319 physically realizable MMs for training. Using the same NN model and architecture, the training process took 4,495.27 seconds. 

Tab. \ref{tab:all_model_metrics} shows the values of the performance metrics for the \textbf{Model II} trained on combined dataset.
Notably, the training performance in \(R^2\) for \(M_{41}\) has increased compared to the corresponding value for \textbf{Model I} (see Tab \ref{tab:model_metrics}).
\begin{table}[H]
\caption{\textbf{Model II}: Performance Metrics}
\label{tab:all_model_metrics}
\begin{tabular}{lllll}
\hline
\textbf{Metric} & \textit{M$_{41}$} & \textit{M$_{42}$} & \textit{M$_{43}$} & \textit{M$_{44}$} \\
\hline
MSE & 0.000030 & 0.000124 & 0.000137 & 0.000193 \\
RMSE & 0.005475 & 0.011138 & 0.011725 & 0.013903 \\
MAE & 0.003003 & 0.006179 & 0.006502 & 0.007533 \\
\textbf{$R^2$} & 0.738724 & 0.918492 & 0.930410 & 0.998285 \\

\hline
\end{tabular}
\end{table}
Using the MM images of the same isolated sample that were recorded with the IMP1, the \textbf{Model II} was tested to evaluate the 
robustness in the performance metrics (see Tab. \ref{tab:all_model_sample24}). The performance scores did not show significant improvement but still show the 
values comparable to initial robustness testing scores from \textbf{Model I}. 

\begin{table}[H]
\caption{\textbf{Model II}: Robustness Testing using the IMP1 data}
\label{tab:all_model_sample24}
\begin{tabular}{lllll}
\hline
\textbf{Metric} & \textit{M$_{41}$} & \textit{M$_{42}$} & \textit{M$_{43}$} & \textit{M$_{44}$} \\
\hline
MSE & 0.000025 & 0.000108 & 0.000190 & 0.000837 \\
RMSE & 0.004985 & 0.010404 & 0.013794 & 0.028929 \\
MAE & 0.003667 & 0.007488 & 0.010137 & 0.016683 \\
\textbf{$R^2$} & 0.393400 & 0.932663 & 0.888064 & 0.940946 \\
SSIM & 0.883913 & 0.937512 & 0.954937 & 0.992077 \\
\hline
\end{tabular}
\end{table}

Tab. \ref{tab:metrics_first_dataset} presents the performance scores from \textbf{Model II} tested on MM images of a skin sample recorded with the IMP2. As expected, the performance scores show significant improvement compared to \textbf{Model I}, which can be explained by introduction of IMP2 data into the training. Overall, the performance scores indicate 
good performance results with low error scores and large SSIM values.
\begin{table}[H]
\caption{\textbf{Model II}: Robustness Testing using the IMP2 data}
\label{tab:metrics_first_dataset}
\begin{tabular}{lllll}
\hline
\textbf{Metric} & \textit{M$_{41}$} & \textit{M$_{42}$} & \textit{M$_{43}$} & \textit{M$_{44}$} \\
\hline
MSE & 0.000013 & 0.000052 & 0.000054 & 0.000027 \\
RMSE & 0.003556 & 0.007224 & 0.007331 & 0.005161 \\
MAE & 0.001511 & 0.002361 & 0.002663 & 0.001918 \\
\textbf{$R^2$} & 0.879348 & 0.851476 & 0.805960 & 0.999812 \\
SSIM & 0.954935 & 0.984081 & 0.976308 & 0.999726 \\
\hline
\end{tabular}
\end{table}

The ultimate test on the accuracy of the reconstruction of the last row of MM with NN model consists in calculating of the diagnostic maps of the depolarization $\Delta$ (dimensionless), diattenuation D (dimensionless), linear retardance R (radians), and orientation of the optical axis $\Psi$ (radians) by applying Lu-Chipman decomposition pixel-wise to the original and reconstructed MM images. 

\begin{figure}[!h]
\centerline{\includegraphics[width=.56\columnwidth]{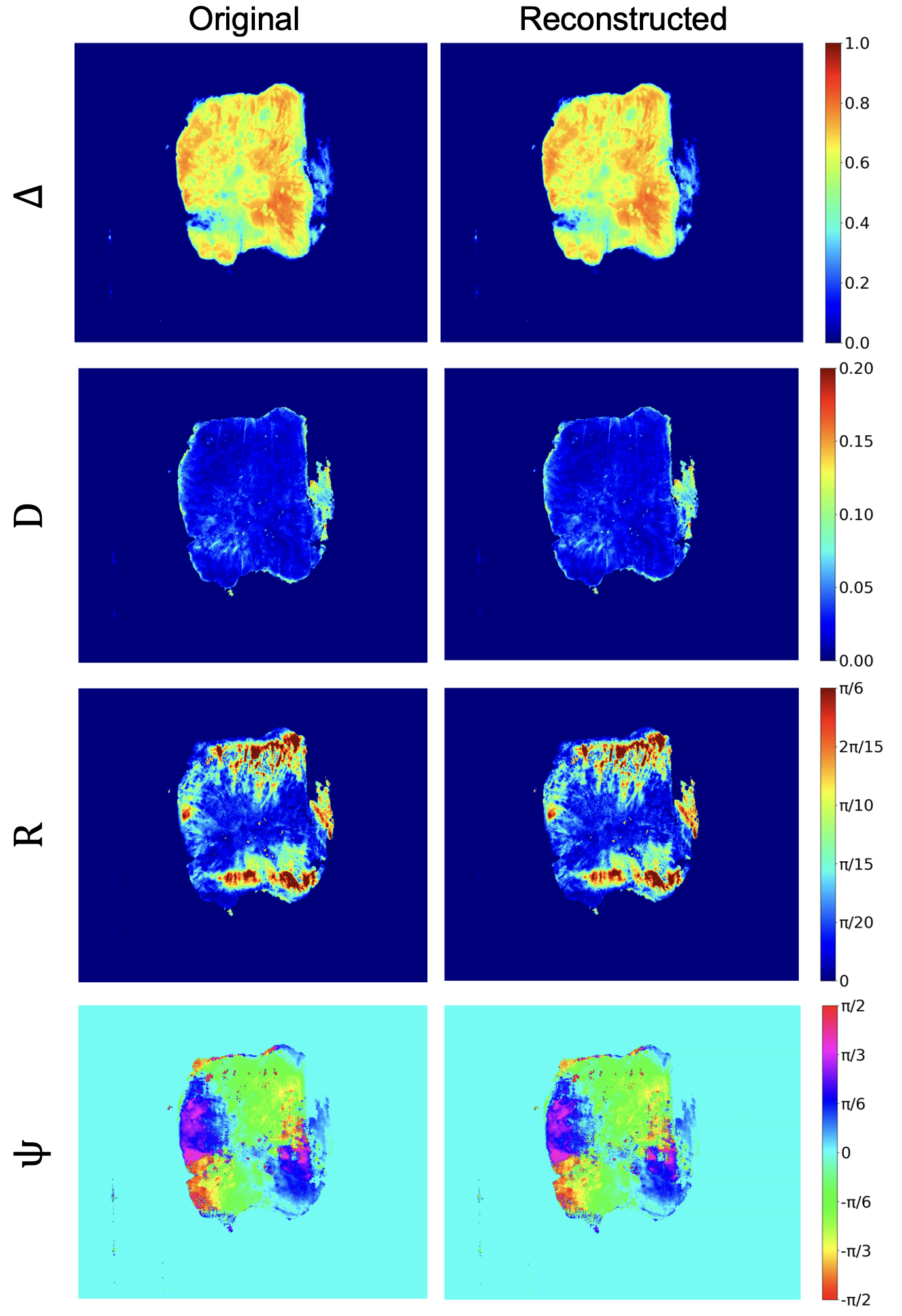}}
\caption{Polarimetric maps of depolarization $\Delta$ (first row), diattenuation D (second row), linear retardance R (third row) and orientation angle of the optical axis $\Psi$ (last row) obtained from the original (left column) and reconstructed (right column) MM images by applying Lu-Chipman decomposition pixel-wise. }
\label{fig:LC_combined}
\end{figure}

An example of such maps calculated from the MM images of excised cervical tissue measured with the IMP1 is shown in Fig. \ref{fig:LC_combined}. 
As expected, the cervical tissue is strongly depolarizing with the border zones demonstrating the values of linear retardance up to $\pi/6$. 
Because of close to normal incidence 
of probing light the diattenuation values are close to zero \cite{DoFP2, novikova2022complete}, except of the specimen edges, where the angle of incidence deviates from normal. 

\begin{figure}[H]
\includegraphics[width=.5\columnwidth]{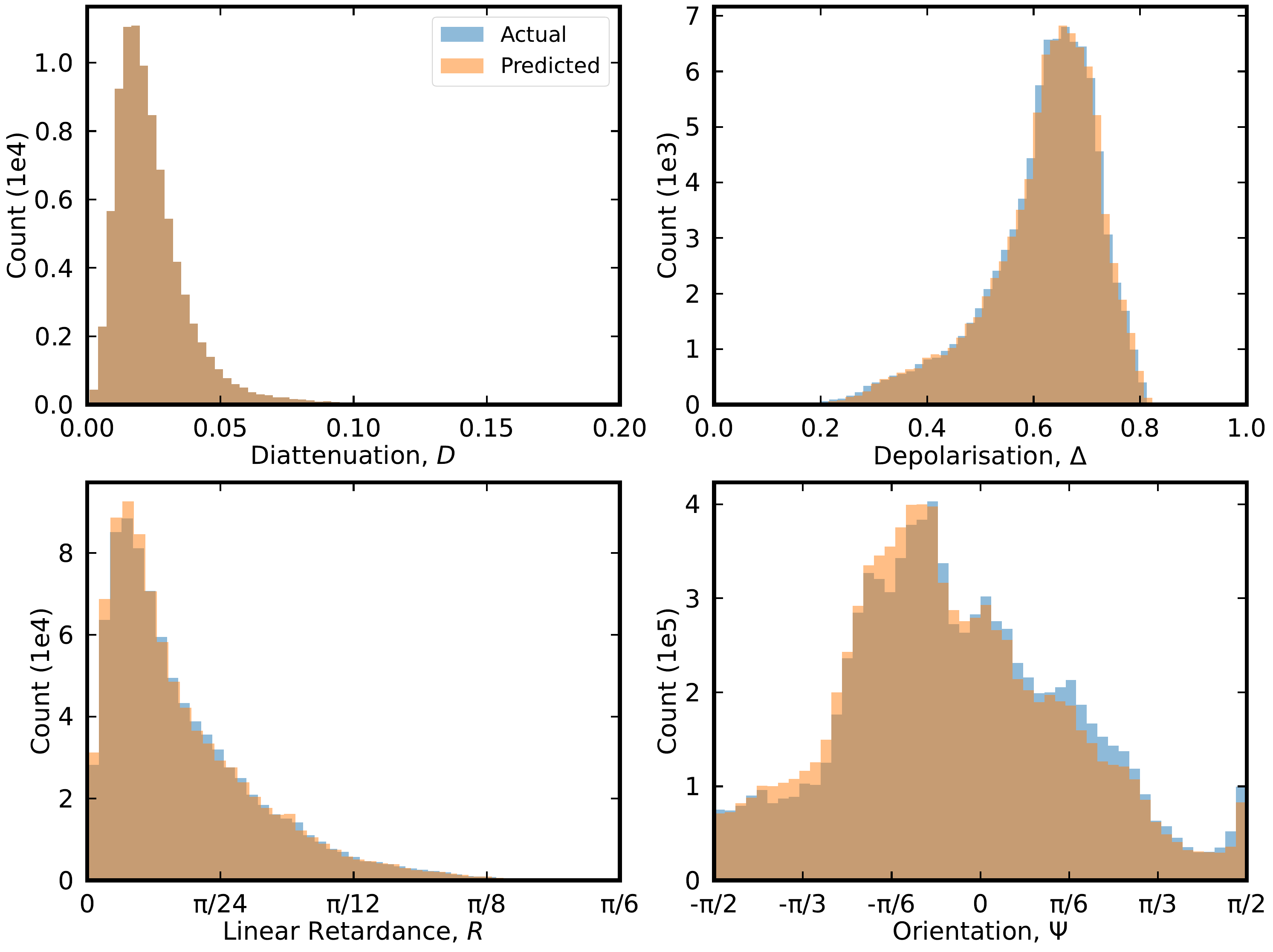}
\caption{Histograms of 
polarimetric parameters D, R, $\Delta$ and $\Psi$ calculated from the original
and reconstructed MMs (see 
Fig. \ref{fig:LC_combined}). Zero values corresponding to the filtered out pixels are omitted.}
\label{fig:histogram_LC}
\end{figure}

As shown in Fig. \ref{fig:histogram_LC}, the histograms of the polarimetric parameters D, R, $\Delta$ and $\Psi$ calculated from the original and reconstructed MMs
only demonstrate small difference.

Tab.~\ref{tab:LuChipman_metric} shows the values of
the performance metrics from \textbf{Model II} for polarimetric parameters calculated with Lu-Chipman decomposition of the original and reconstructed MMs of cervical specimen (see Fig.~\ref{fig:LC_combined}). 
%
\begin{table}[H]
\caption{\textbf{Model II}: Robustness Testing using Lu-Chipman decomposition of IMP1 data}
\label{tab:LuChipman_metric}
\begin{tabular}{lllll}
\hline
\textbf{Metric} & \textbf{D} & \textbf{$\Delta$} & \textbf{R} & \textbf{$\Psi$} \\
\hline
MSE & 0.0 & 0.000017 & 0.000008 & 0.004636 \\
RMSE & 0.0 & 0.004071 & 0.002833 & 0.068089 \\
MAE & 0.0 & 0.001271 & 0.000696 & 0.024537 \\
\textbf{$R^2$} & 1.0 & 0.999721 & 0.999929 & 0.925642 \\
SSIM & 1.0 & 0.996232 & 0.982431 & 0.940401 \\
\hline
\end{tabular}
\end{table}
As expected, the SSIM and $R^2$ values for the reconstructed and original polarimetric maps of diattenuation D are equal to 1, all error metrics are equal to 0, because this parameter is calculated from the elements of the first row of MM which are identical for the original and reconstructed matrices. 

The SSIM and $R^2$ values are very close to 1 
for the depolarization $\Delta$, 
linear retardance R and 
orientation angle of the optical axis $\Psi$,
thus, demonstrating a high level of similarity between the two sets of images. 
The error scores have low values for the depolarization $\Delta$, linear retardance R and orientation angle of the optical axis $\Psi$. 

However, it is 
still to explore whether these small differences in polarimetric maps may impact the accuracy of clinical diagnosis, e. g. the delineation of the pathological zones.
%
\section{Conclusion}
\label{sect:conclusion}
This study has demonstrated the potential of machine learning techniques to 
reconstruct the complete 4$\times$4 MM from its partial measurements. Utilizing a sequential NN model, the last row of 3$\times$4 MM was generated, thereby completing the matrix.  
Through further training with different tissue types, measured at different configurations and wavelengths, our results show last row prediction still works well. 
%
Without any parameter adjustments, we 
reconstructed the images of the last row of MM for uterine cervix, brain, skin and colon tissues. In our studies the reconstruction was done sequentially pixel by pixel. The execution time of the trained NN model was about 300 microseconds for a single pixel. Performing the reconstruction of the missing elements of the fourth row of MM in parallel for all image pixels would ensure the execution time of the algorithm is compatible with the partial MM data acquisition at video rate. 

However, this approach's limitation 
is the black-box nature of deep learning models, which hinders our understanding of how the model forms these relationships. Consequently, 
to extend the developed model for other types of tissues, first, one will need to use the complete MM images collected on these specific tissue types for the model training process.

We also showed that the quality of input data (i. e., the physical realizability of all MMs in a
dataset) is very important for the good performance of the NN model. 
%
The model's robust performance, evidenced by low error metrics and high \(R^2\) values across various tissue samples, suggests that the machine learning approach can effectively be used in combination with the existing methods of decomposition of depolarizing 4$\times$4 MMs (e. g. Lu-Chipman decomposition) for the post-processing of partial polarimetric data. This approach paves the way for real-time \emph{in-vivo} applications of 3$\times$4 Mueller polarimetry and its translation to clinical practice.

\section*{Author contributions statement}
S.~C. developed and applied
the sequential neural network model, performed 
data post-processing and statistical analysis. T.~H. 
implemented the hybrid calibration method of IMP2. 
S.~C, T.~H collected MM images of tissue thin sections with IMP2, wrote the manuscript. O. R-N. built 
IMP2 and with J.-C.~V. developed LabView-based control interface. A.~A., J.~C. R.-R. developed Matlab-based control interface, T. H. and T.~L. adapted it to 
IMP2. J. V., A. P. collected MM images of cervical specimens with 
IMP1. Thin tissue sections
were prepared and provided by G. P. (brain) and T. G. (skin, colon). T. N. designed the studies. A. D., H. M., and T. N. supervised the studies, interpreted data, and edited the manuscript. All authors have read and approved the manuscript.
\section*{Acknowledgment} This work was supported in part by the funds from Swiss National Science Foundation (Sinergia grant HORAO \#CRSII5\_205904);  
Campus France PHC “Dumont D’Urville” (POLANNs grant \#43261UE
); Royal Society of New Zealand (Contract E3430); Campus France PHC “Rila” (AURORA grant \#48152NJ, grant \#KP-06-Rila/3-15.12.2021); Campus France PHC "Osmose" (PHAETHON grant \#50590ZG); COST Action \#CA21159 PhoBioS; Institut National du Cancer and Cancéropôle, France (PAIR Gynéco contract \#2012-1-GYN-01-EP-1), NSF STROBE grant \#1828705.

T. H. acknowledges China Scholarship Council funding of his 6 months
research stay at LPICM, Ecole polytechnique.
%
T.~G. acknowledges the kind cooperation of Prof. P.~Troyanova and Dr. I. Terziev from the University Hospital ”Tsaritsa Yoanna – ISUL”, Sofia, Bulgaria, in the preparation of 
skin and colon tissue sections.
T. N. acknowledges the contribution of Dr. J. Rehbinder from the ICube laboratory, University of Strasbourg, France, to cervical data collection. 
T. N. and T. G. dedicate this paper to Prof. Ekaterina Borisova's memory. 

\section*{Disclosures} The authors declare no conflicts of interest.


\section{Appendix}
\label{App:1}
\begin{table}[H]
\caption{Summary of Neural Network Hyperparameters and Architecture}
\label{tab:nn_hyperparameters}
\footnotesize 
\begin{tabular}{ll} 
\hline
\textbf{Parameter/Architecture} & \textbf{Specification} \\
\hline 
Input Layer Shape & (12,) \\
First Dense Layer Units & 320 (ReLU activation) \\
Second Dense Layer Units & 224 (ReLU activation) \\
Third Dense Layer Units & 320 (ReLU activation) \\
Output Layer Units & 4 \\
Optimizer & Adam \\
Learning Rate & 0.0001 \\
Loss Function & Mean Squared Error \\
Batch Size & 32 \\
Epochs & 20 \\
Total Parameters & 448,046 \\
Trainable Parameters & 149,348 \\
Non-trainable Parameters & 0 \\
\hline
\end{tabular}
\end{table}


\newpage
\section{Supplementary materials}
%
Our study employed two different instruments for dataset collection to enhance data variability and develop a more robust neural network model. 
\subsection{IMP1 system}
\begin{figure}[h]
    \begin{center}    \includegraphics[width=0.4\columnwidth]{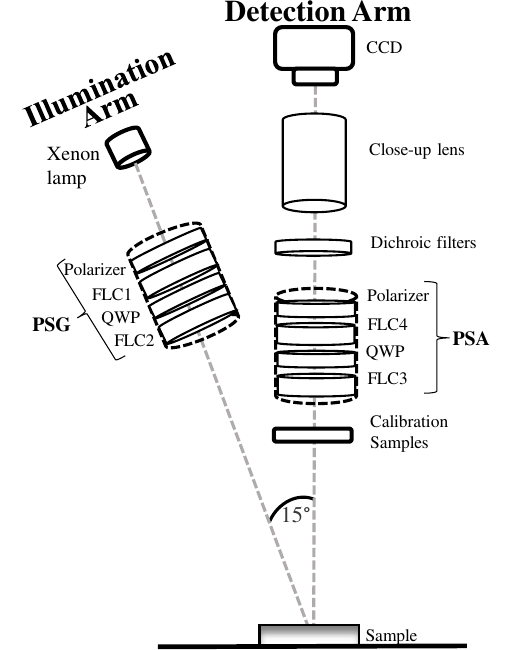}
    \end{center}
    \caption{Schematic of the optical layout of the IMP1}\label{fig:IMP1}
\end{figure}

The first instrument is a wide-field imaging Mueller polarimeter (IMP1) operating in reflection geometry, as shown in Fig.\ref{fig:IMP1}. This ferroelectric liquid crystal (FLC)-based system operates in the visible wavelengths ranging from 450 nm to 700 nm \cite{Lindberg2019,  RodriguezNunez2021, Robinson2023}. The instrument is composed of an incoherent white light source (Xenon lamp), followed by the Polarization State Generator (PSG) for modulation of the incident light beam illuminating the sample. After hitting the sample light is collected in the detection arm that includes the Polarization State Analyzer (PSA), followed by the spectral filters, the zoom lens, and the CCD-sensor (Stingray F080B ASG). 

The PSG consists of a linear polarizer with the transmission axis at 0$^{\circ}$ relative to the laboratory reference frame, followed by the FLC1, a quarter-wave plate (QWP), and a second FLC2. The PSA consists of the same components as the PSG but arranged in the reverse order. The IMP1 records sequentially 16 intensity images (4$\times$4 matrix \textbf{I}) of a sample in a few seconds. The calibration matrices \textbf{A} and \textbf{W} of the PSA and PSG, respectively, are obtained with the eigenvalue calibration method \cite{Compain}. Finally, the 4$\times$4 MM of a sample is calculated as \textbf{M} = \textbf{A}$^{-1}$\textbf{I}\textbf{W}$^{-1}$. %
\subsection{IMP2 system}
\begin{figure}[h]
    \begin{center}    \includegraphics[width=.7\columnwidth]{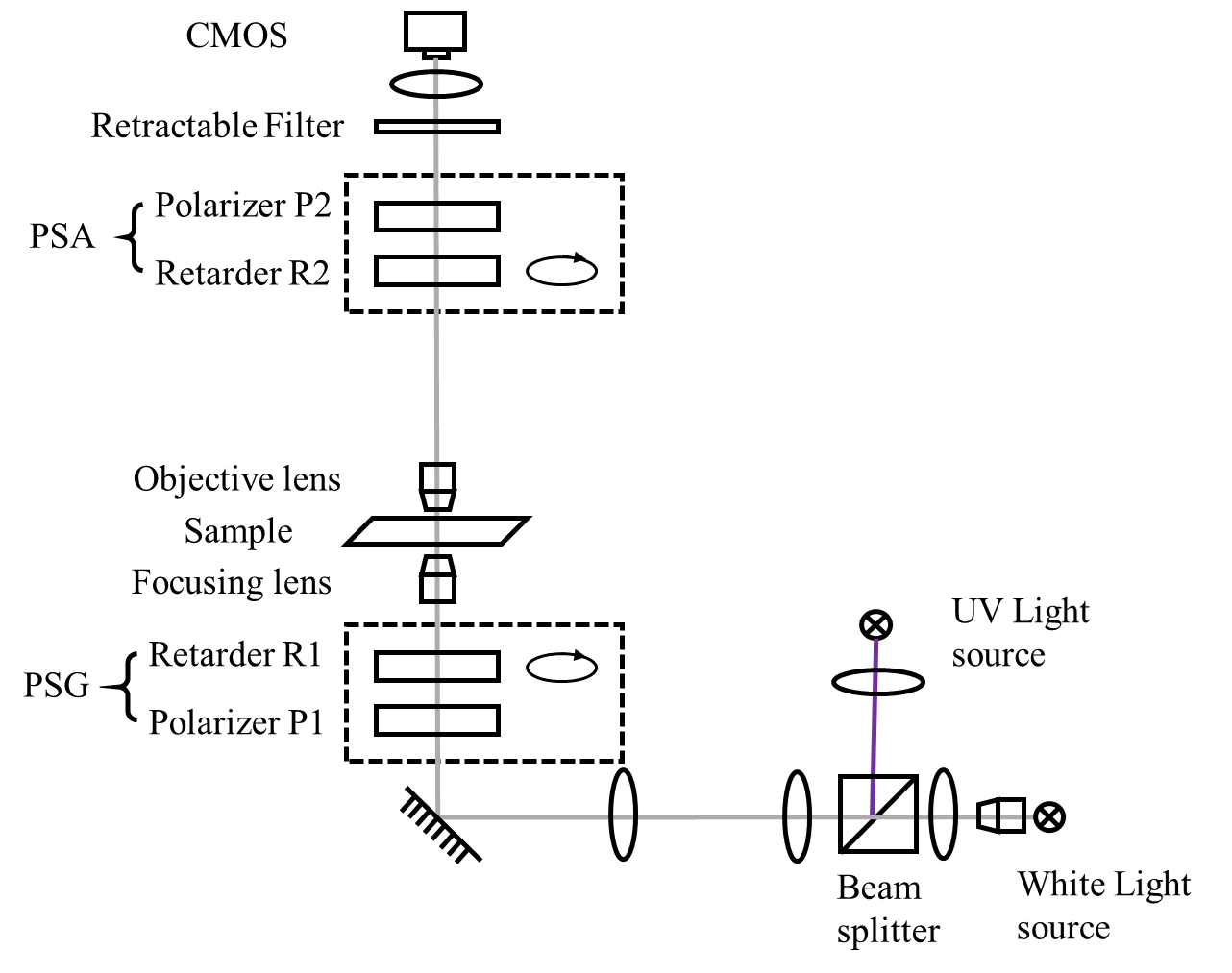}
    \end{center}
    \caption{Schematic of the optical layout of the IMP2.}\label{fig:IMP2}
\end{figure}

The second part of the dataset was 
acquired using
a custom-built transmission MM microscope (IMP2), shown in Fig.\ref{fig:IMP2}. The instrument features two continuous light sources: a UV-A LED (385 nm, 1650 mW, Thorlabs, France, M385LP1) and a visible light LED (6500 K, 2350 mW, Thorlabs, France, MCWHLP1). 
The polarization state generator (PSG) comprises a fixed Glan-Taylor polarizer (15.0 mm CA, Thorlabs, France) and a rotating achromatic quarter-wave plate (Thorlabs, France). The system also includes two microscope objectives (4× and 20× Nikon Plan Fluor Imaging Objectives, Nikon, Japan) and a polarization state analyzer (PSA) consisting of the same optical components as the PSG, but arranged in a reverse order. The detection is performed using a CMOS camera (2.1 Megapixel Monochrome Compact sCMOS Camera, USB 3.0, Thorlabs, France).

Unpolarized light from the LED source is modulated by the PSG in the illumination arm before passing through the first microscope objective, which focuses the light and increases its intensity on the sample. Light transmitted and scattered by the sample then passes through the second microscope objective, followed by the PSA and the CMOS camera in the detection arm. A retractable bandpass filter is placed in front of the CMOS camera for measurements in the visible wavelength range. The horizontal linear polarizer P1 in the PSG is oriented at 90$^\circ$ relative to the second linear polarizer P2 in the PSA, and both are aligned parallel to the initial orientation of the fast optical axes of the quarter-wave plates R1 and R2 in the PSG and PSA, respectively. During the measurement of the Mueller matrix (MM), both R1 and R2 sequentially rotate to four pre-selected orientation angles\cite{Sabatke2000, Chang2016} of the optical axis: ±51.7$^\circ$, ±15.1$^\circ$. This rotation results in 16 intensity images captured by the CMOS camera. The Mueller matrix \textbf{M} of a sample can be reconstructed as \textbf{M} = \textbf{A}$^{-1}$\textbf{I}\textbf{G}$^{-1}$. where \(\textbf{G}\) and \(\textbf{A}\) stand for the instrument matrix of PSG and PSA, respectively. 
\(\textbf{I}\) stands for 4$\times$4 detected raw intensity matrix 
Hybrid calibration method is implemented to calculate the matrices \(\textbf{G}\) and \(\textbf{A}\)
\cite{huang2023hybrid}.



\end{document}